CBN 14-01                                                                                                           March 10, 2014

# RESULTS ON FIELD MEASUREMENTS IN A FLAT POLE MAGNET WITH THE CURRENT CARING SHEETS

*Alexander Mikhailichenko*

*Abstract.* The results of measurements with a gradient magnet, arranged with the help of current sheets are represented and analyzed.

## INTRODUCTION

In [1] it was suggested to make gradient magnets for CESR upgrade with the help of techniques which includes thin current sheets covering the flat poles of existing dipole magnets, Fig.1. The field gradient required for such upgrade migrated from the initial values 0.2 *kG/cm* to ~2 *kG/cm* for ~6.5*GeV* beam ([2]-[4]), and the current required finally arrived to the 6-7kA per sheet. Although for CESR upgrade it looks now as too high, it might be interesting for some others projects, so we decided to finalize the experimental research by this brief publication.

Let us mention here, that similar technique was used for creating gradient in a bending magnet [5], where "the plane corrector bars were mounted between the vacuum chamber and poles". The current in bars was ~250 A[1]. Correcting gradient in a magnet with conductors attached to the poles of dipole magnet is a techniques used in synchrotron B3-M at BINP, Novosibirsk since 1965, [6].

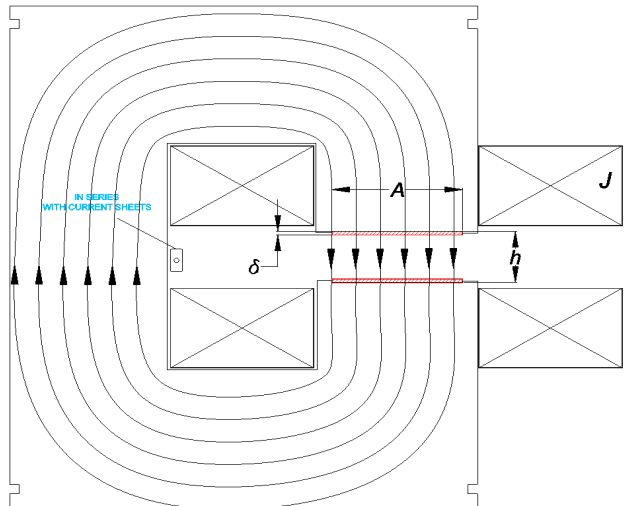

**Figure 1**. The concept of gradient arrangements in the magnet. Thin sheet as a conductors (marked red) are running all along the magnet poles and carry the current ~5 kA each [1].

---

[1] Unfortunately publication [5] does not describe details; there is no indication that they used compensation conductor for keeping the central field value steady, while the current in bars is changed.

The operational gradient which could be obtained with technique described in [1] is

$$G = \frac{dB(x)}{dx} \cong \frac{2I_0}{A \cdot h} \text{ (SI units)}, \quad G = \frac{dB(x)}{dx} \cong \frac{0.8\pi I_0}{A \cdot h} \text{ (practical units)} \quad (1)$$

where $I_0$ stands for the total current value in each sheet, $A$ is a sheet width, $h$ is the gap between the poles of magnet. Substitute here for example $I_0$=5000A, $h$=5 cm, $A$=10 cm one can obtain

$$G \cong \frac{0.8\pi I_0}{A \cdot h} = \frac{0.8\pi \, 5000}{10 \cdot 5} \cong 250 G/cm = 25 kG/m \cong 2.5 T/m, \quad (2)$$

i.e. a moderate one. If the current sheet has a thickness $\delta$=3mm, then the current density comes to a value $\frac{I_0}{A \cdot \delta} = \frac{5000}{100 \cdot 3} \cong 17 A/mm^2$, which requires cooling.

The cooling arranged by soldering the water carrying copper tube(s) to the sheet. As the conducting tube redistributes the current flow, it should be attached to the plate at the sides running along the magnet pole. Anyway, calculations carried with numerical codes can take this into account easily. We used extraction of some material of sheet so the effective current density in a sheet and the tube wall, integrated over the vertical coordinate at any radius remains the same.

Following this concept described in [1], a practical model was implemented in a spare CESR magnet of reduced length, see Fig 2. This was done for investigation on how practical this concept is for CESR and others projects.

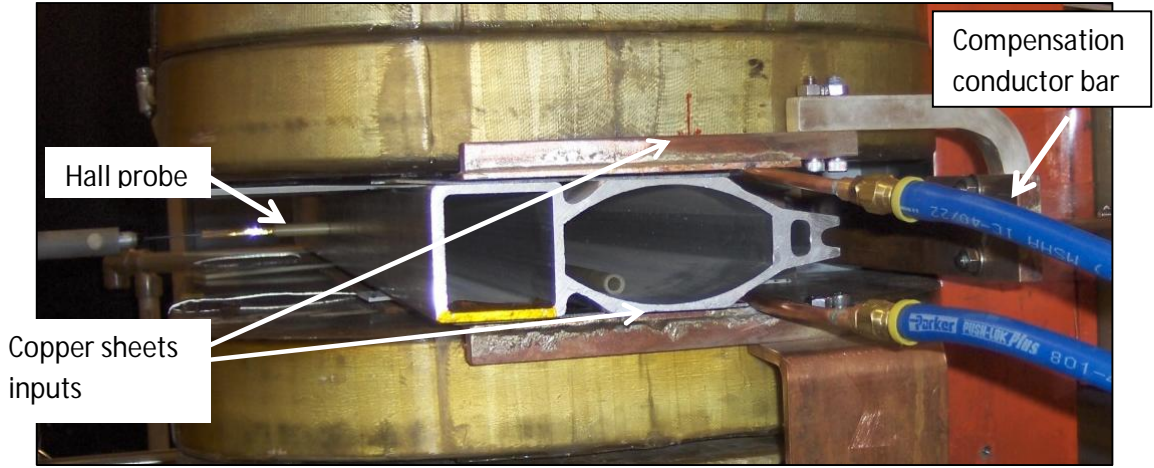

**Figure 2.** Two insulated~3mm-thick current sheets inserted in the gaps between the vacuum chamber and the upper and lower poles. The sheets are running along the magnet poles. Hall probe is moving across the chamber in the middle through the holes drilled in the walls of chamber. Water cooling tubes (blue ones) feed the water cooling tubes brazed to the sheets at the side edges.

Fortunately, the vacuum chamber of CESR has vertical dimension less than the pole gap allowing allocation of 3 mm-thick sheets wrapped by a Kapton tape between the chamber and poles at the top and bottom of chamber. The magnet is ~1.5 meter long; the section of vacuum chamber in this model has about the same length. The walls of chamber drilled through in a horizontal direction for possibility to move the Hall probe across the aperture in the middle of magnet.

**MEASUREMENTS OF THE FIELD DISTRIBUTION IN A TRANSVERSE DIRECTION**

Setup for measurements of field elevation in a transverse direction is represented in Fig.3 below. Here the Hall probe could be moved by stepping motor across the aperture, through the walls of chamber. The current in a driving magnet coil delivered by PS, 500A typically, while an additional PS, which feeds the sheets (connected in series), and was able to run up to 3 *kA* (only).

During operation, stabilization of current was not the class of operation at CESR, just the one delivered by the factory made PS (see Fig.16). Results of field readings for different current excited in a sheet are represented in Fig.4.

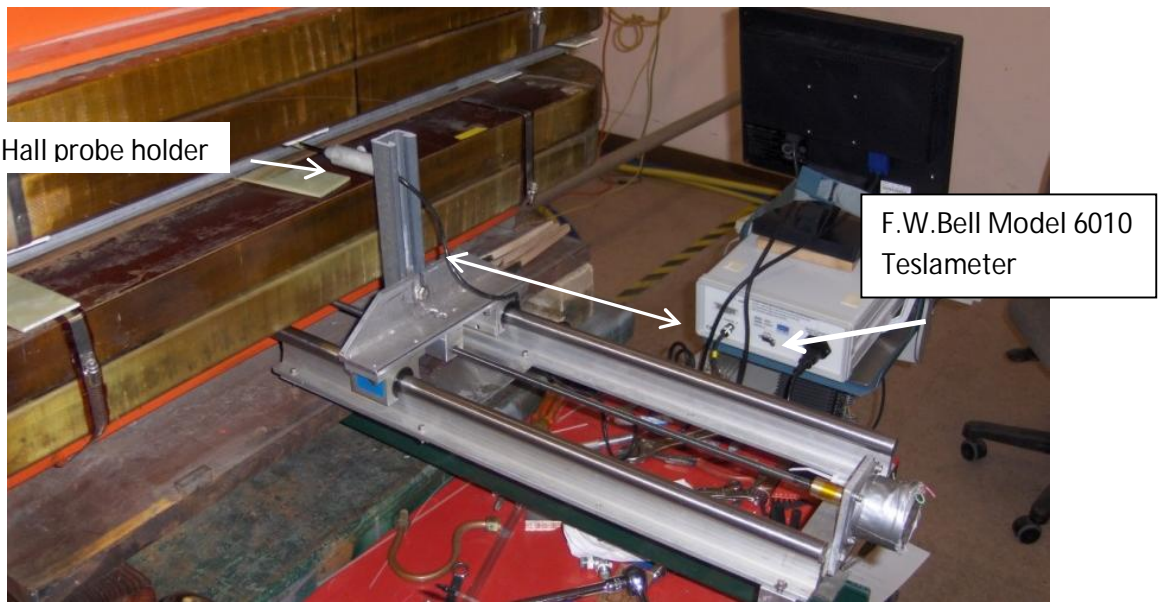

**Figure 3.** Cartridge with Hall probe could be moved in a transverse direction through the holes drilled in a vacuum chamber.

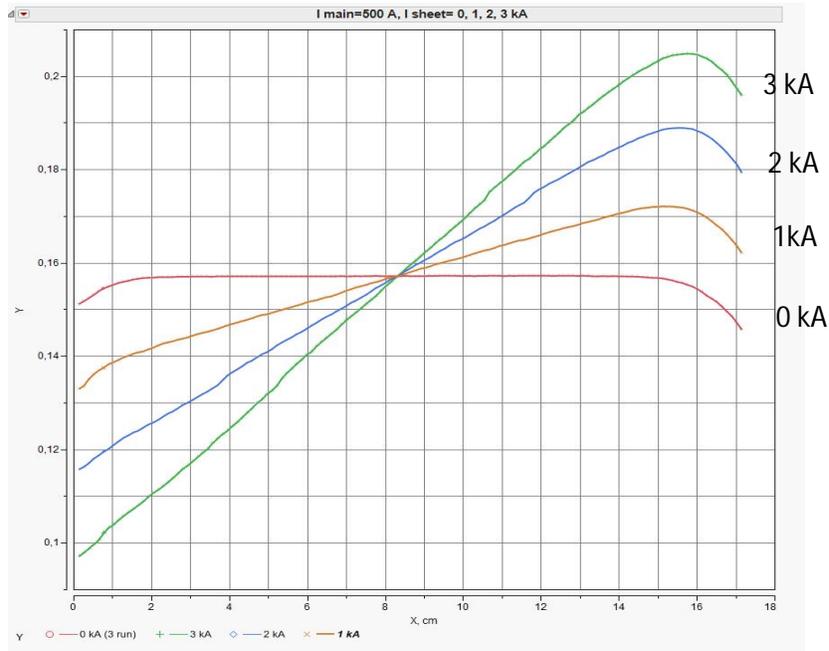

**Figure 4.** Field elevation across the aperture for different currents in a current sheet.

In Fig.5, a part of previous figure is zoomed for better visualization of central point. For ideal field stabilization the lines in Fig.5 should be straight ones, however for the real demonstrated stability of PS they are sporadically curved at the central region. Even so, one could see that the center point holds its position with accuracy better, than 0.1mm.

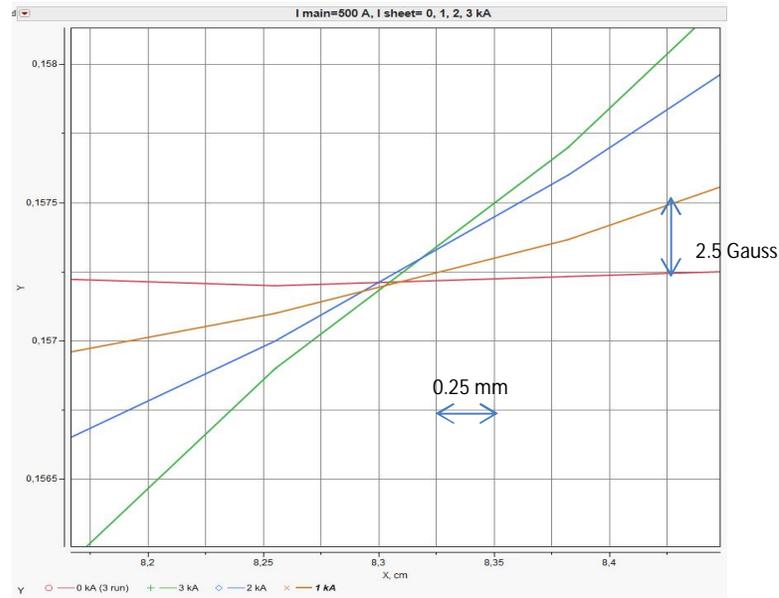

**Figure 5.** Field elevation across aperture for different currents in a current sheet; zoomed.

So we concluded, that the method of the field shift compensation at central region by additional current bar [1] is a satisfactory one indeed.

**MEASUREMENTS OF FIELD DISTRIBUTION IN A LONGITUDINAL DIRECTION**

For measuremnts of field elevation in a longitudinal direction we used a cartridge, driven by a stepping motor, which was used for measurements of SC wiggler magnet [7], see Fig. 6 and Fig 7. The Hall probe could be positioned at different transverse offsets and for each position, the longitudinal run delivered longitudinal distribution of vertical field, Fig.8 and Fig. 9. Stepping motor was appointed for the longitudinal positioning of cartridge.

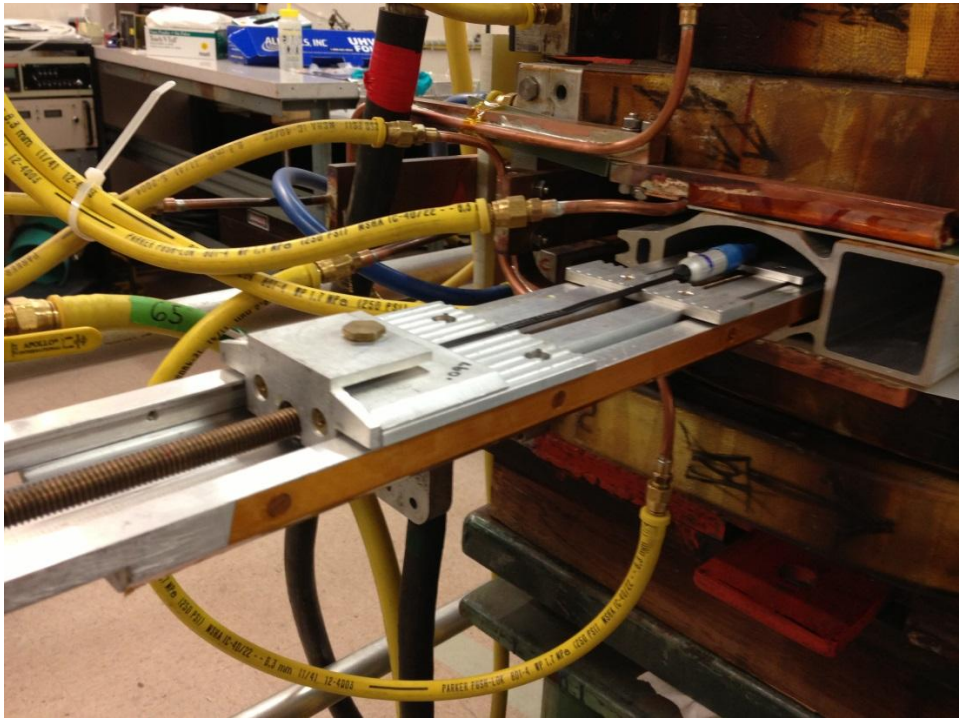

**Figure 6.** Hall probe cartridge exit at the South end. Stepping motor is at the left; out of figure.

We would like to mention here, that the midplane level of cartridge is ~4*mm* higher, than the midplane of magnet; this is due to the difference in a chamber shape in regular magnets of CESR and in the wigglers. Modification of cartridge required substantial modification, what was recognized as inadequate to the task effort.

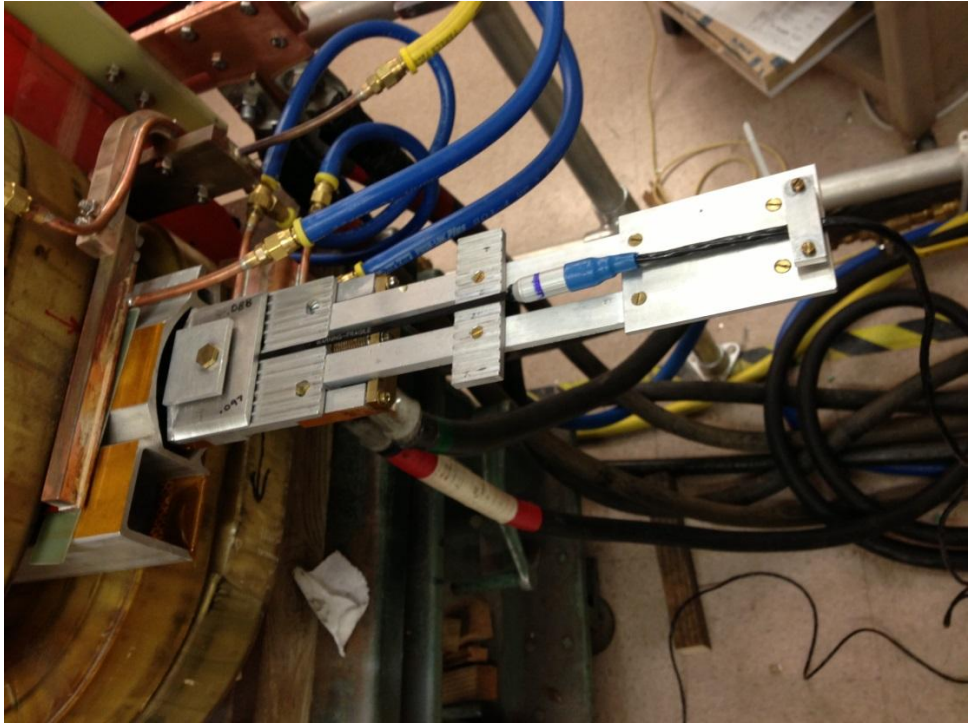
**Figure 7.** Hall probe exit at the opposite side.

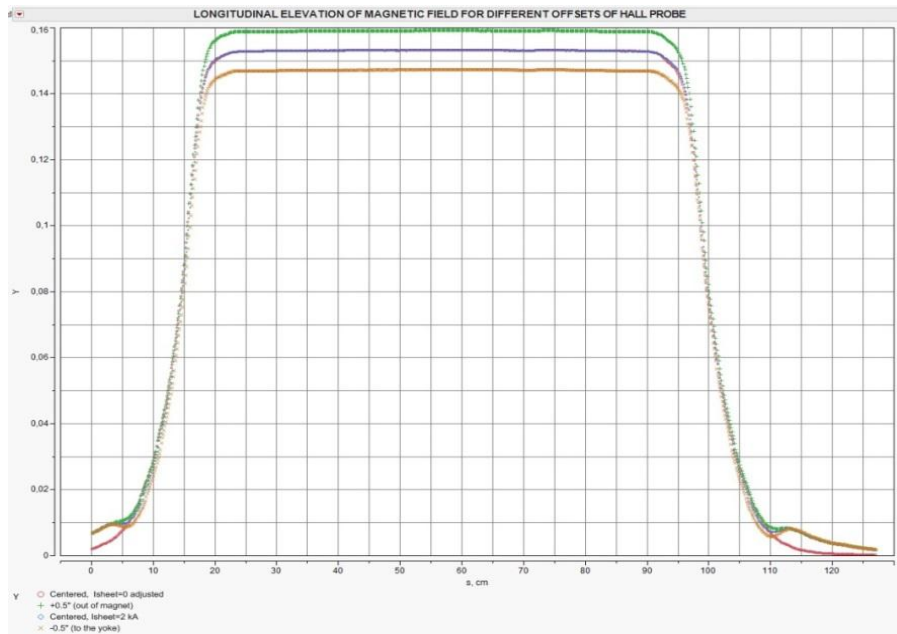
**Figure 8.** Longitudinal elevation of magnetic field for different offsets of Hall probe. I$_{sheet}$=2*kA*.

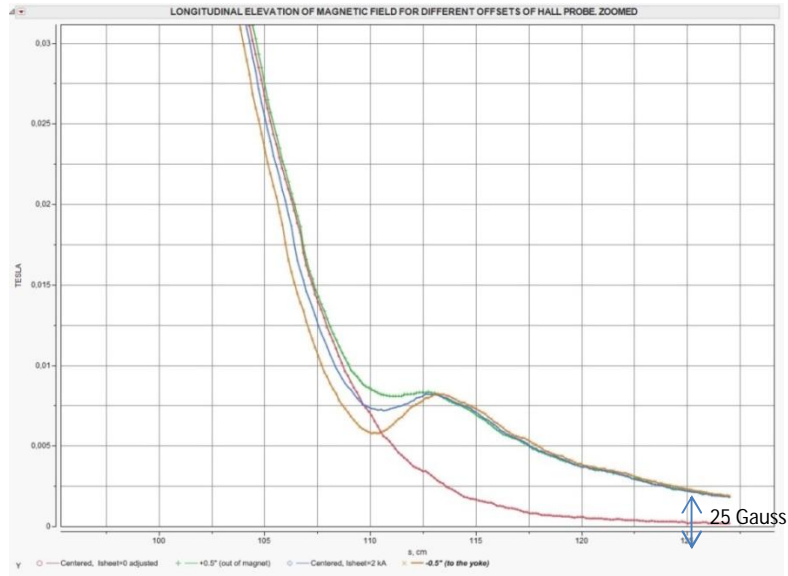

**Figure 9.** Longitudinal elevation of magnetic field for different offsets of Hall probe; Fig.8 zoomed.

One can see that there is some kind of bump at the fringe region, which was explained by asymmetric input of current in a sheet. Offset of medial plane of magnet and cartridge mentioned above gives input in the bumps also.

## MODIFICATION OF INPUT CONDUCTORS

Conductors were modified according to the following diagram, Fig. 10:

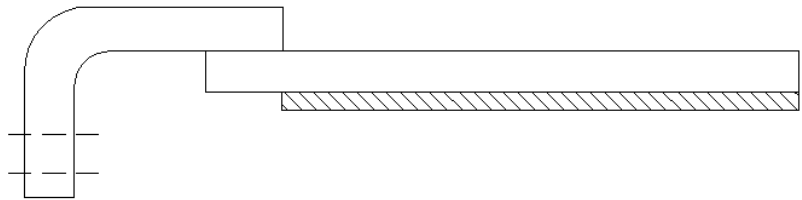

*a) Initial input contact configuration*

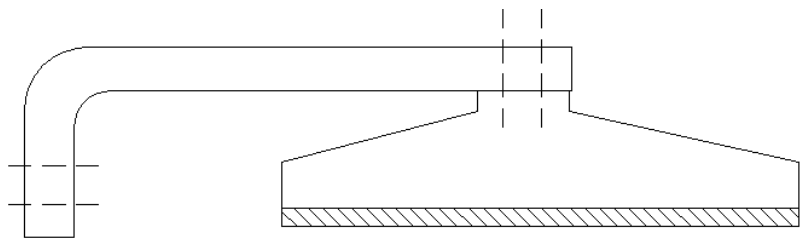

*b) Symmetrized input of current into the current sheet*

**Figure 10.** Modification of input conductors.

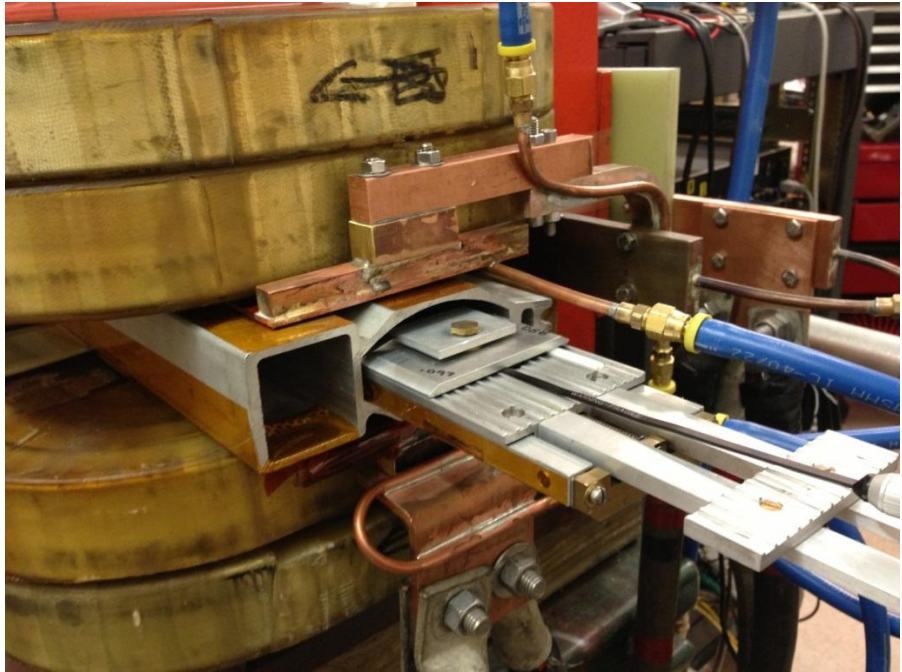

**Figure 11.** View to the modified end input.

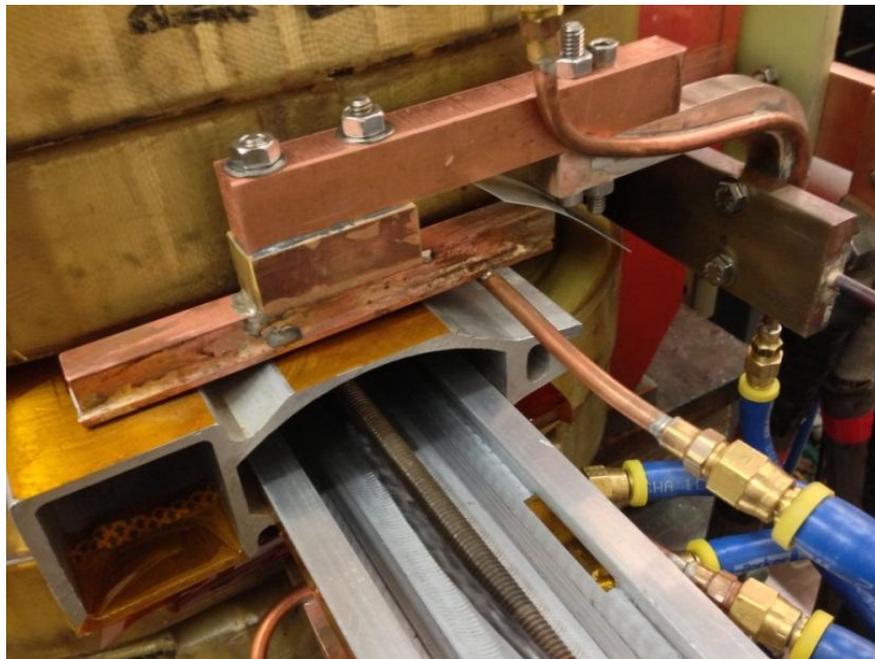

**Figure 12.** Opposite entrance. Lower current sheet input has the same configuration.

Although the input is still different from what is desirable, see Fig. 15 below (from [1]), the bumps are practically disappeared. This confirmed the recognized source of such behavior.

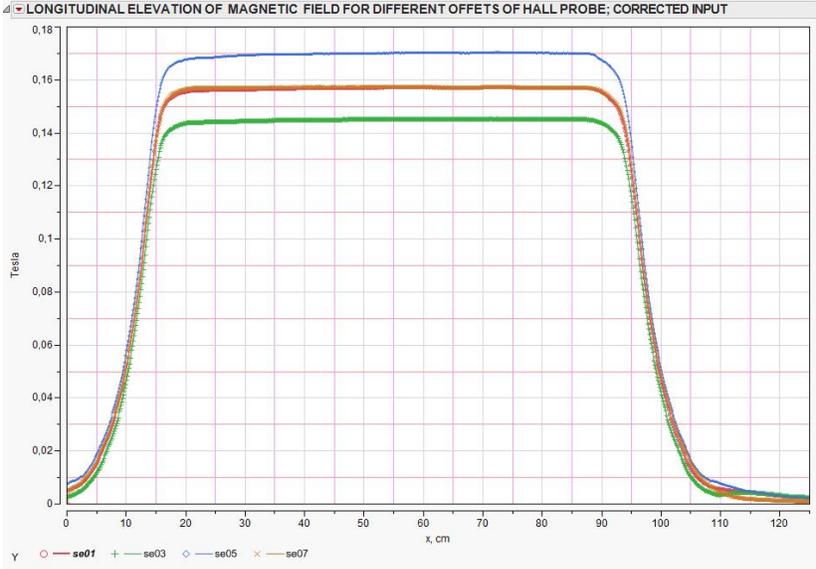

**Figure 13.** Longitudinal elevation of magnetic field for different offsets of Hall probe. $I_{sheet}$=2kA. Modified input (compare with Fig. 8).

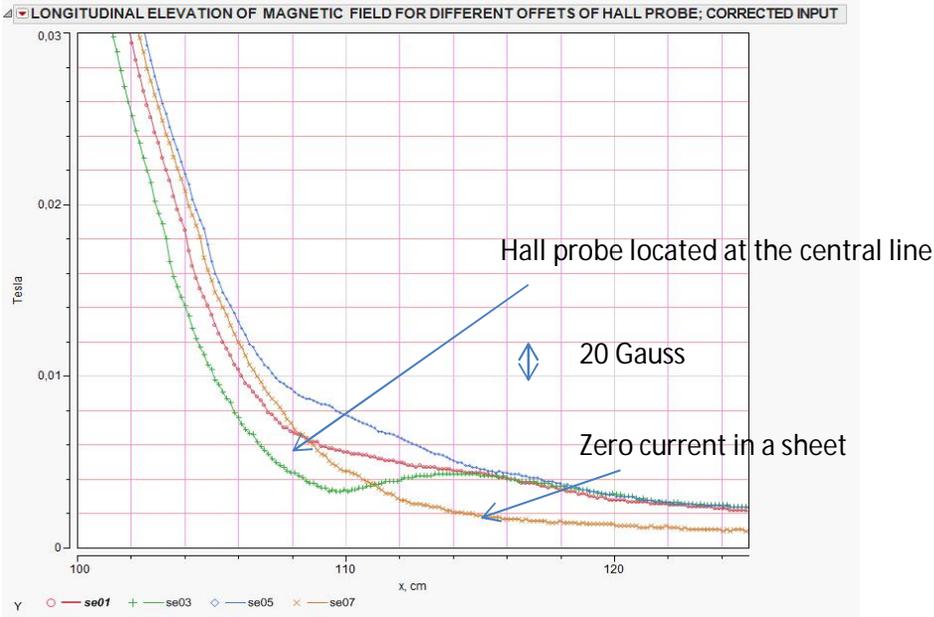

**Figure 14.** Longitudinal elevation of magnetic field for different offsets of Hall probe; Fig.13 zoomed.

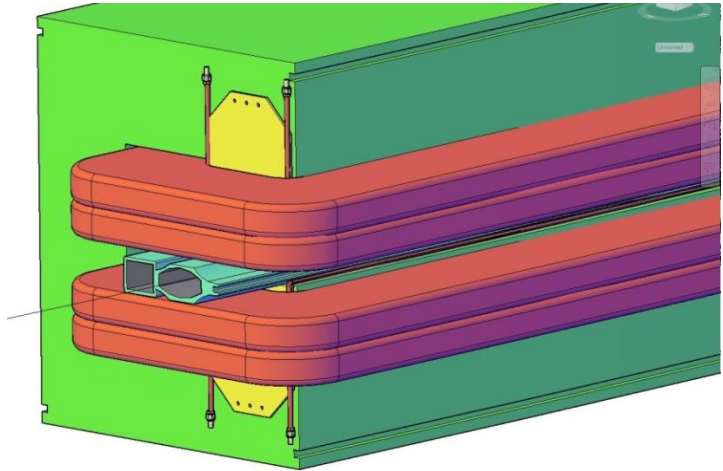

**Figure 15.** Desirable allocation of the current sheet inputs [1]. Current sheets (yellow) are running out of the fringe region.

So the conclusion of these exercises is obvious: the method of gradient generation in a flat-pole magnet with a sheet-like conductor might be acceptable in some projects, may be not for CESR upgrade however.

## RECOMMENDED MODIFICATION OF PS

Below we would like to describe some possible modification of power supply, suitable for feeding the magnets with current-sheet gradient corrector.

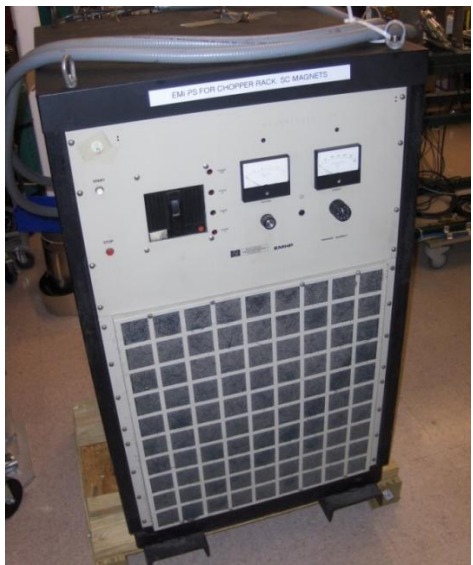

**Figure 16.** Old fashion PS used for feeding the current sheets; able to generate 3 kA in sheets.

Newer PS supplies use the high frequency chopped voltage to feed the loard through the transformer, rectifying diodes and cables running to the load, Fig. 17, left.
For reduction of losses in cables it is naturally to move the transformer and diodes closer to the load, Fig.17, right.
The concept of high-frequency transformer with low stray field is shown in Fig.18 and Fig.19.

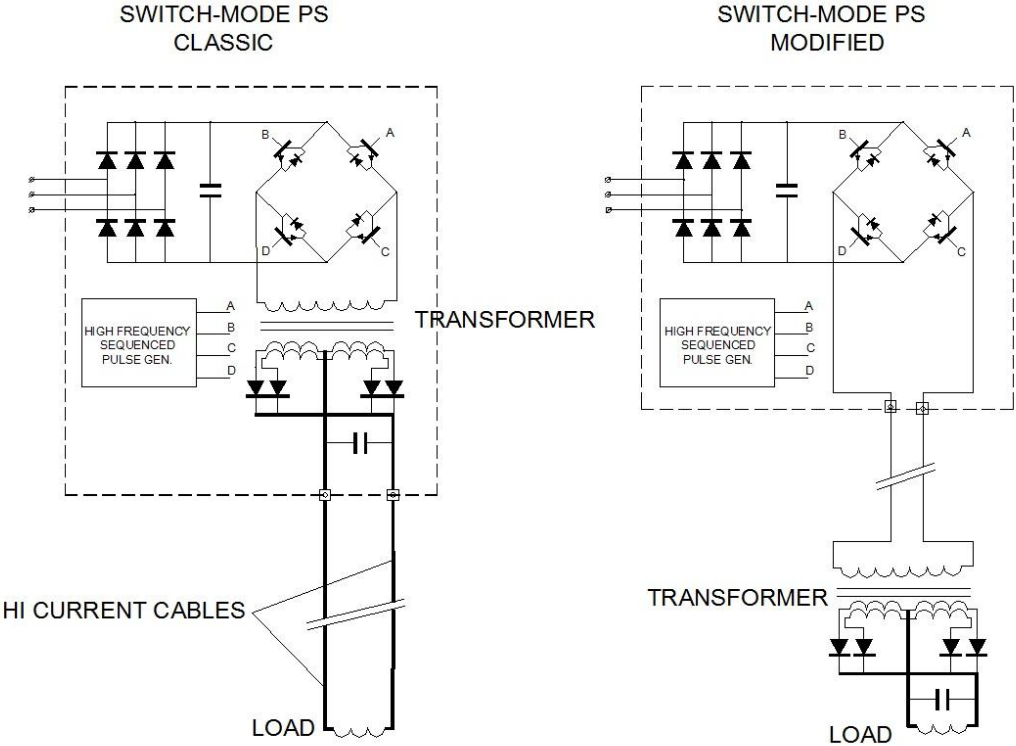

**Figure 17.** Transformer with rectifying diodes moved towards the load.

The cross section of transformer is shown in Fig. 17. Upper sketch represents the variant with outer location of rectifying diodes, the lower one shows an engineering variant with inner location of diodes. For high chopping frequency $f$ the transformer dimensions are reduced ~$1/f$ , so the allocation of transformer closer to the load is not a problem..

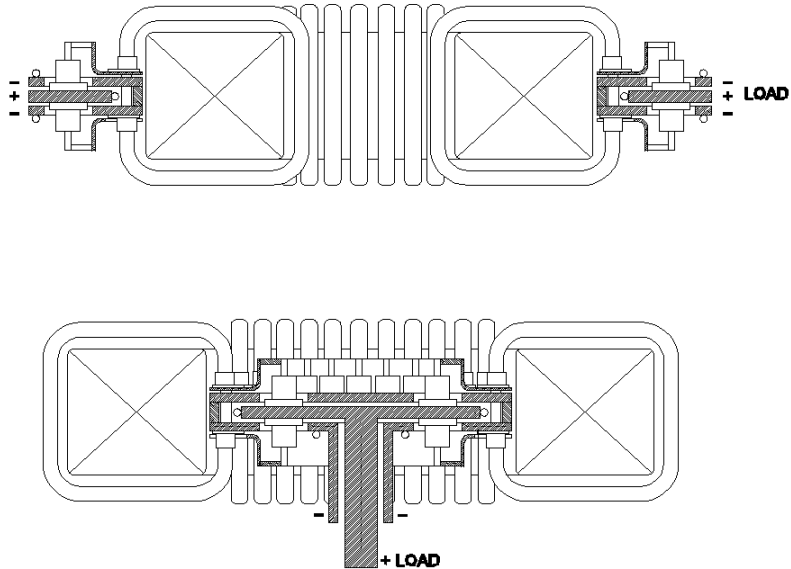

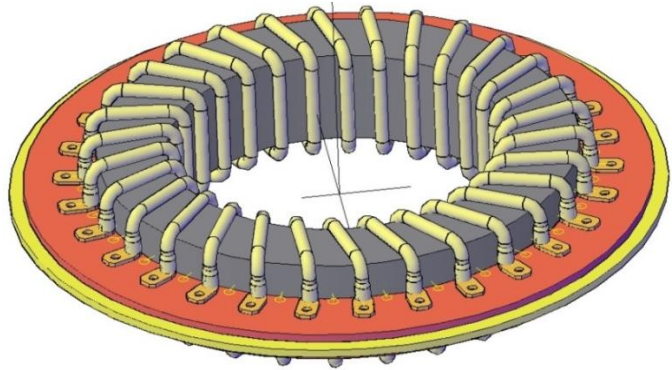

**Figure 18.** Cross-section of transformer.

**Figure 19.** The concept of a cable transformer; 3D sketch.

## CONCLUSION

Magnet with regulated gradient implies a useful instrument for the beam tuning and operation. Obvious disadvantage of the method - high feeding current- could be balanced by its simplicity for some projects. Even a fraction of gradient controllable variation is always useful. We would like to say that correction of gradient by many current conductors located at the magnet pole has a long history, which begins, practically, from the earlier days of accelerator physics (Cornell Synchrotron, B3-M at Budker INP, Novosibirsk).

Gradient estimated for CESR upgrade, (2) is $G\sim 250G/cm$; however for smaller aperture magnet, say 2.5$cm$ pole gap and ~5 $cm$ pole width the gradient reaches ~1$kG/cm$. Five mm-thick sheets could carry current 5$kA$ with current density~20$A/mm^2$, so the heat could be evacuated by water flow easily, see Fig.20.

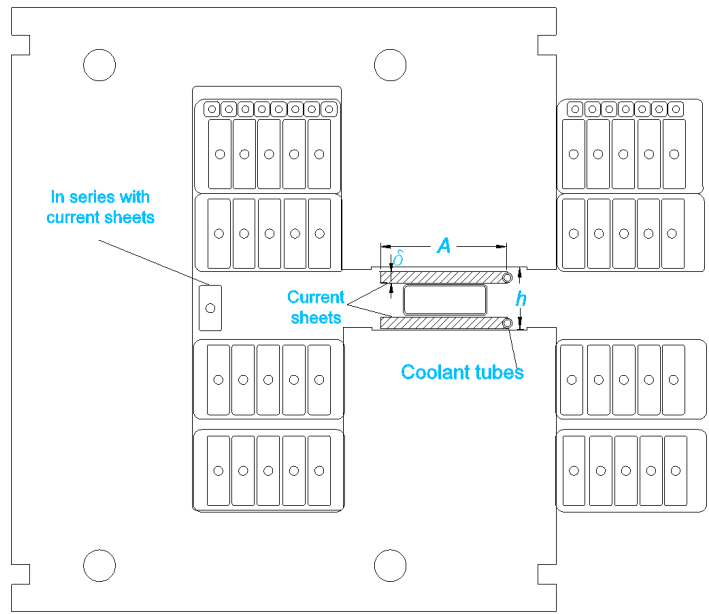

**Figure 20.** To the ultimate value of gradient.

This type of gradient correction could be recommended for ILC damping ring and for many other projects as well.

In conclusion author thanks William Trask for his help in assembling this setup.